\documentclass[11pt]{article}
\usepackage{amssymb}
\usepackage{amsfonts}
\usepackage{amsbsy}
\usepackage{amsmath}
\usepackage{amsthm}
\usepackage{amscd}
\usepackage{amstext}
\usepackage[pdftex]{hyperref}

\def\ben{\begin{equation}}
\def\een{\end{equation}}

    \let\L=\Lambda
 \let\P=\Phi

\def\be{\begin{equation}}
\def\ee{\end{equation}}
\def\beq{\begin{equation}}
\def\eeq{\end{equation}}
\def\ba{\begin{array}}
\def\ea{\end{array}}

\def\dalemb#1#2{{\vbox{\hrule height .#2pt
       \hbox{\vrule width.#2pt height#1pt \kern#1pt
               \vrule width.#2pt}
       \hrule height.#2pt}}}

\newcommand{\bea}{\begin{eqnarray}}
\newcommand{\eea}{\end{eqnarray}}

\def\pt{\partial_t}
\def\pr{\partial_r}
\def\pf{\partial_\phi}

\def\FTheta{{\bf{\Theta}}}

\def\FTheta{{\bf{\Theta}}}
\def\L0vac{{\cal L}_0^{vac}}
\def\P0vac{{\cal P}_0^{vac}}
\def\ffL0{{\cal L}_0}
\def\ffP0{{\cal P}_0}
\def\SWald{S_{\text{Wald}}}

\begin{document}

\begin{center}

{ \LARGE {\bf BTZ Black Holes and Flat Space Cosmologies in Higher Derivative Theories}}

\vspace{1.2cm}

 C\'eline Zwikel 

\vspace{0.9cm}

%{\it $^\sharp$ ULB \\}
\centerline{\sl Universit\'e Libre de Bruxelles, Physique Th\'eorique et Math\'ematique 
% and International Solvay Institutes
}
\centerline{\sl Campus Plaine C.P. 231, B-1050 Bruxelles, Belgium}
\smallskip

%\vspace{0.5cm}
%
%{\it $^*$ Theory Group, SLAC National Accelerator Laboratory \\
%Menlo Park, CA 94025, USA \\}

\vspace{1.6cm}

{\tt czwikel@ulb.ac.be} \\

\vspace{1.6cm}

\end{center}

\begin{abstract}
We consider BTZ black holes %with two sets of boundary conditions, Brown-Henneaux and Compère-Song-Strominger, 
and flat space cosmologies in generic higher derivative gravity theories in 2+1 dimensions. 
Our goal is to prove the match between the bulk Iyer-Wald entropy and the field theory entropy for various symmetry algebras (CFT$_2$, Warped CFT$_2$, BMS$_3$). We also discuss phase transitions in higher curvature theories, and argue that, in the flat case, %the study of stability shows 
there is strictly speaking no phase transition in the grand canonical ensemble.
%, except in the case of flat chiral gravity. 
%This result differs from previous works in the literature. 
%as opposed to 
%and describe the Hawking-Page transition. 
%The BTZs are studied equipped with the Brown-Henneaux and also with the Compère-Song-Strominger boundary conditions. Their asymptotic symmetry groups is the two Virasoro and the semi-direct product of a centrally extended Virasoro algebra and an affine $u(1)$ Kac-Moody algebra, respectively. 
%The latter has as asymptotic group the $bms_3$ algebra. 
%derived in a putative warped conformal field theory.  In this paper, we show that this entropy-matching continues to hold for the most general higher derivative theories of gravity.
%and $3d$ flat space cosmologies 
%We give also an explicit expressions of exacts and central charges for any gravity theory. 
%For BTZ, the charges are a renormalization of general relativity charges and terms with Riemann derivatives in the Lagrangian doesn't contribute at all. The entropies matching follows immediately. Moreover, the phase diagram of the Hawking-Page transition is identical at the general relativity case. 
%The charges of flat space cosmologies are always equal to the general relativity charges. Trivially, the charges properties derived in general relativity still holds in the generic case. 
%We argue that there is no phase transition by studying the local stability, as opposed at what was done in 
\end{abstract}

\pagebreak
\setcounter{page}{1}

\tableofcontents

\pagebreak

\section{Introduction}
%The seminal work of Brown-Henneaux have shown that the asymptotic symmetry group of AdS$_3$ was the conformal group in two dimensions. 

%In a 2d CFT, the Cardy formula counts the asymptotic density of states. Strominger \cite{Strominger:1997eq} has shown it is reproduced exactly by the Bekenstein-Hawking entropy of BTZ black hole \cite{Banados:1992gq,Banados:1992wn}.  It constitutes a non trivial check of the AdS/CFT correspondence conjectured by Maldacena \cite{Maldacena:1997re}. 

In a 2D conformal field theory (CFT), the Cardy formula counts the asymptotic density of states \cite{Cardy:1986ie}. Strominger \cite{Strominger:1997eq}, building on Brown-Henneaux's seminal work \cite{Brown:1986nw} relating asymptotically AdS$_3$ spacetimes to a 2D CFT, has shown it exactly accounted for BTZ black holes entropy \cite{Banados:1992gq,Banados:1992wn}. This result constitutes one of the milestones of the close relationship between AdS gravity and conformal field theories \cite{Maldacena:1997re}. 

%The AdS/CFT correspondence conjectured by Maldacena \cite{Maldacena:1997re} have passed all the checks up to now and there are no clues of the contrary.
%In 3D gravity, one of the most famous check concerns the entropy of the BTZ black hole \cite{Banados:1992gq,Banados:1992wn}. Strominger \cite{Strominger:1997eq} has shown it is exactly reproduced by the Cardy formula, counting the asymptotic density of states in a 2D CFT. 

In the spirit of a UV complete theory of gravity, higher curvature terms are expected to appear as corrections to its low energy limit, general relativity. Therefore, it is interesting to get to know if the entropy matching still holds for such theories. 
%In general, we will consider a diffeormorphism invariant theory of gravity (with or without derivative of the Riemann). 
%From the string theory point of view, the higher curvature terms correspond to corrections, suppressed by the inverse of the string tension $\alpha'$, to its low energy limit, general relativity.
First, Saida and Soda \cite{Saida:1999ec} have proven that the matching is preserved in any theory of gravity without derivative of the Riemann tensor in the Lagrangian. 
They performed a frame transformation by defining a new metric bringing the original higher curvature frame into an Einstein frame - the new metric obeys to the Einstein-Hilbert action - and auxiliary fields, which do not contribute to the gravitational entropy. The new metric is always proportional to the BTZ metric. Consequently, all charges are just multiplied by that proportionality constant. 
%They conjecture that it should be true for a theory even with derivative of the Riemann.
Later, Kraus and Larsen \cite{Kraus:2005vz} have shown that the matching occurs in any theory of gravity for solutions with a near horizon AdS$_3\times X$ geometry. Their derivation uses the special asymptotic behaviour of AdS spacetimes for Brown-Henneaux boundary conditions. This method is not directly transposable to others cases of interest, in particular the boundary conditions introduced in \cite{Compere:2013bya,Barnich:2006av}.

%It is important to note that is a powerful result and stronger that what we could expect. Indeed, general relativity can be viewed as a low energy limit of string theory.  \cel{Ca ne ma pas l'air juste. }
One particular term in 3D gravity is the Chern-Simons term, bringing a gravitational anomaly. 
 For BTZ black holes, its effect has already been studied in \cite{Tachikawa:2006sz,Hotta:2008yq,Kraus:2005zm,Bouchareb:2007yx,Solodukhin:2005ah} and therefore, it will not be considered in this work. 
% In this work, in general we don't consider this term expect when it is explicitly mention.

%Before describing the content of this paper, we would like to stress that we are not including the Topologically Massive Gravity (TMG) \cite{Deser:1981wh,Deser:1982vy} case, expect when it is explicitly mention.  Indeed it is particular because it has a gravitational anomaly and for the solutions considered here, it has been already studied separately in the literature. 

%you that here we will treat separately the Topologically Massive Gravity (TMG) case which is particular because it has a gravitational anomaly and for the solutions considered here, it has been already studied in the literature. 
%It is sufficient to add its contribution at the end because the definitions of charges have a linear dependence in the Lagrangian. 

In this paper, we first recover the results of \cite{Saida:1999ec,Kraus:2005vz}, i.e. the matching between the statistical and gravitational entropy and the expressions of charges for BTZ black holes with Brown-Henneaux boundary conditions \cite{Brown:1986nw}, using another method: the covariant phase space formalism. We are inspired by \cite{Azeyanagi:2009wf} where the authors studied the entropy matching for 4D extremal Kerr black holes. We use this formalism throughout this paper.

%has expected to be corrected by higher curvature terms
%the higher curvature terms are corrections in the expansion in $\alpha'$, the string tension to the low energy limit, namely general relativity. 
% general relativity can be viewed as a low energy limit of string theory.  
%At high energies, the Einstein-Hilbert action is expected to be corrected by higher curvature terms suppressed by a dimensionful scale. Typically in string theories, the coupling would correspond to the inverse string tension.

%from both side of the duality 

%Indeed, adding higher curvature term can change drastically the situation.

%One of the goal of that paper is to prove the conjecture made by Saida and Soda \cite{Saida:1999ec}: the gravitational entropy matches with the Cardy formula for higher curvature theories with derivatives of the Riemann.
%Note that the particular case TMG was done in \cite{}. 

%To compute the entropy in the gravity side, they use the Iyer-Wald formula \cite{Iyer:1994ys} and in CFT side, they compute the central charge in the Hamiltonian formalism. 
%First, new boundary conditions have been recently found for BTZ black holes by Comp\`ere, Song and Strominger \cite{Compere:2013bya}. 
Secondly, we consider  BTZ black holes equipped with the recently new boundary conditions found by Comp\`ere, Song and Strominger \cite{Compere:2013bya}. The asymptotic symmetry group is generated by a $U(1)$ Kac-Moody-Virasoro algebra, suggesting a description in terms of a warped conformal field theory \cite{Detournay:2012pc}.
%The dual CFT is there strongly believed to be a warped conformal field theory \cite{Detournay:2012pc}. 
%Initially, those new types of field theories arose in the description of warped AdS$_3$ spacetimes.
% \footnote{For warped spacetimes, the generalization of the entropy matching to any gravity theory has been studied in a companion paper \cite{Detournay:2016gao}.  }.
%In that paper, it is the other ensemble and BTZ is included in that story by taking the warping parameter to one. It worthily to say that here the direct use of symmetries
There, the counting of the asymptotic density of states can be made using an analog of the Cardy formula, called the warped Cardy formula.
Nevertheless, its form depends of the considered ensemble, closely related to the choice of coordinates. Here, we consider the coordinates of \cite{Compere:2013bya} referring to the quadratic ensemble (see section 4 and 5.3 of \cite{Detournay:2012pc}) where the warped Cardy formula takes the form of a Cardy formula. 
%\cel{In \cite{Detournay:2010rh}, the show the matching between the entropies for TMG. }
In this paper, we show that the gravitational entropy of BTZ can be reproduced by a warped Cardy formula for any gravity theory. 
%The particular case was already proved in \cite{Detournay:2012pc}. 
%The Saida-Soda method still is applicable but we will miss the theory with derivatives of the Riemann. We use another technique presented below.

Another interesting class of spacetimes in three dimensions are the flat space cosmologies \cite{Cornalba:2002fi}. These are locally flat, their asymptotic symmetry group is given by BMS$_3$ \cite{Barnich:2006av}, and can be obtained as a certain limit of BTZ black holes \cite{Cornalba:2002fi,Barnich:2012aw}. 
Their entropy is also reproduced by the Cardy-like formula for BMS$_3$ \cite{Barnich:2012xq,Bagchi:2012xr}.
In this paper, we derive the charges in any diffeomorphism invariant theories and also prove the entropy matching. 

%\cel{Why are they interesting}
%Here, we generalize them to any higher curvature theories of gravity.  

%We use the method developed in \cite{Azeyanagi:2009wf},i.e. covariant phase space formalism. Therein they show the matching between the statical and gravitational entropy for 4d extremal Kerr black holes. 
%The choice of that formalism and the use of symmetries, 
%However in our cases, BTZ black holes and flat space cosmologies, 
%in addition to show the matching of entropies for any gravity theories, allow us to give an explicit expressions for all charges and moreover show analytically that there are no effects on the charges from the higher derivative corrections. 
%However in our cases, BTZ black holes and flat space cosmologies, in addition to show the matching of entropies for any gravity theories, the symmetries allow us to give an explicit expressions for all charges and moreover show analytically that the higher derivative corrections vanished.
%(which was not the case in \cite{Azeyanagi:2009wf} ). 
%One should mention the particular case of TMG where there the charges are modified \cite{Bagchi:2013lma,}. 

%These computations will also lead us to have an explicit expressions for the charges. 
%\cel{Showing by the same time the integrability of the cha}

Finally, we consider phase transitions. 
%through the global stability as well as the local stability. 
In the case of AdS$_3$ geometries, it was shown that a phase transition occurs between the thermal AdS$_3$ spacetime and the BTZ black hole (see \cite{hawking1982,Maloney:2007ud,Detournay:2015ysa} and references therein). 
%The extensions to some other theories was done in \cite{Detournay:2015ysa} \cel{OTHER???}. 
We show that the higher curvature corrections do not modify the phase diagram. 
% here that the general case reduces to the behaviour in general relativity. 
For flat space cosmologies, it was argued in \cite{Bagchi:2013lma} that there exists also a phase transition between them and the hot flat spaces in the grand canonical ensemble. 
%Here, we show in both cases, that the phase transition behaviour in any theory reduces to the 
Nevertheless, their analysis of local thermodynamic stability needs to be refined and leads to the conclusion that these spacetimes are not locally stable in the grand canonical ensemble for any theories without gravitational anomalies. 
 %So in that context, the study global stability does not make sense since it is a comparison between local minima of the Gibbs free energy. 
 The study of phase transition at thermodynamic equilibrium requires the comparison between minima of free energies. Therefore, in that context, we exclude the possibility of a phase transition. 
% So there is no point to study the global stability which is a comparison between local minima of the Gibbs free energy. 
We also analyse the local stability in the particular cases of topologically massive gravity (TMG) \cite{Deser:1981wh,Deser:1982vy} and flat chiral gravity \cite{Bagchi:2012yk}, where only the Chern-Simons term subsists in the Lagrangian. 
%Only the latter allows locally stable flat space cosmologies and there we exhibit a phase transition. 
Only the latter seems to allow locally stable flat space cosmologies and exhibits a phase transition. But a careful analysis of the consequence of the enhancement of symmetries in flat chiral gravity forbids us to consider the thermodynamics of these solutions and thus the phase transition. 

The plan of the paper is the following: in section 2, we define the quantities needed to compute charges. Then, in section 3, we consider the BTZ black holes.
%We first derived the features of an invariant tensor under $SL(2,\mathbb{R})\times SL(2,\mathbb{R})$.
Using the symmetries, we compute the charges and show the entropy matching. After, we study the Hawking-Page transition and present examples of theories where we apply that formalism. In the section 4, we prove the entropy matching for the flat space cosmologies and discuss the phase transition. Finally we conclude.

Note that throughout  this paper, we set Newton's constant to one.

\section{Method and definitions}

In this section, we explicit the expressions of the charges in the covariant phase space formalism without any details. We closely follow \cite{Azeyanagi:2009wf}. 
The most general Lagrangian can be written as \cite{Iyer:1994ys}
\begin{equation}
L=\star f(g_{ab},R_{abcd},\nabla_{e_1}R_{abcd},\nabla_{(e_1} \nabla_{e_2)}R_{abcd},...,\nabla_{(e_1} ...\nabla_{e_n)}R_{abcd}) \, ,
\end{equation} 
or equivalently in terms of auxiliary fields 
\begin{align}
L=&\star f(g_{ab},\mathbb{R}_{abcd},\mathbb{R}_{abcd|e_1},...,\mathbb{R}_{abcd|e_1...e_k})+ Z^{abcd}(R_{abcd}-\mathbb{R}_{abcd})\\
& Z^{abcd|e_1}(\nabla_{e_1}\mathbb{R}_{abcd}-\mathbb{R}_{abcd|e_1})+...+ 
 Z^{abcd|e_1...e_k}(\nabla_{(e_k}\mathbb{R}_{abcd|e_1...e_{k-1})}-\mathbb{R}_{abcd|e_1...e_k})\,.
\end{align}
The auxiliary fields $ Z^{abcd|e_1...e_s}$, $\mathbb{R}_{abcd|e_1...e_s}$ for $1\leq s\leq k$ are totally symmetric in the indices $e_i$ and the symmetrization in $\nabla_{(e_s}\mathbb{R}_{abcd|e_1...e_{s-1)}}$ is only among the $e_i$'s. 
The equations of motion for $\mathbb{R}_{abcd|e_1...e_s}$ and $Z^{abcd|e_1...e_s}$ are
\begin{align}
& \mathbb{R}_{abcd|e_1...e_s}=\nabla_{(e_s}\mathbb{R}_{abcd|e_1...e_{s-1})} \\
& Z^{abcd|e_1...e_s}=\frac{\partial f}{\partial \mathbb{R}_{abcd|e_1...e_s}}-\nabla_{e_{s+1}}Z^{abcd|e_1...e_{s+1}}
\end{align}
where there is no covariant derivative for $s=0$ or $k$.
They can be solved iteratively
\begin{align} \label{auxRabcd}
& \mathbb{R}_{abcd|e_1...e_s}=\nabla_{(e_1}... \nabla_{e_s)}R_{abcd}\\
\label{Zabcd}
& Z^{abcd}=\frac{\delta^{cov}}{\delta^{cov} R_{abcd} }f(g_{ab},R_{abcd},\nabla_{e_1}R_{abcd},...)\,, 
\end{align}
with 
\begin{equation}
\frac{\delta^{cov}}{\delta^{cov} R_{abcd} } = \sum_{i=0}(-1)^i\nabla_{(e_1}\cdots \nabla_{e_i)} \frac{\partial }{\partial\nabla_{(e_1}\cdots \nabla_{e_i)} R_{abcd}}\,.
\end{equation}
%Q_{\xi}_{c_3...c_n}=\left (-Z^{abcd}\nabla_c\xi_d -2 \xi_c\nabla_d Z^{abcd} \right )\epsilon_{abc_3...c_n}
%\end{equation}
The auxiliary field $Z^{abcd}$ allows us to write easily the relevant quantities to compute the charges. In 3D, the Noether charge for a Killing vector $\xi$ is \cite{Azeyanagi:2009wf}
\begin{equation}\label{Noetherchargesdef}
Q_{\xi}=\left ( \left (-Z^{abcd}\nabla_c\xi_d -2 \xi_c\nabla_d Z^{abcd} \right )+
\xi_k A^{kab}\right )\epsilon_{abc}dx^c
 \, ,
\end{equation}
with 
\begin{align}\nonumber
A^{kab}=&-2( Z^{klcd |e_1...e_{s-1}a}\mathbb{R}^b_{lcd|e_1...e_{s-1}}+
Z^{alcd|e_1...e_{s-1} b}\mathbb{R}^k_{lcd|e_1...e_{s-1}}+
Z^{alcd|e_1...e_{s-1} k}\mathbb{R}^b_{lcd|e_1...e_{s-1}}\\\label{corrA}
&+\frac{s-1}2 Z^{lmcd|e_1...e_{s-2} ka}\mathbb{R}^{\phantom{lmcd|}b}_{lmcd|\phantom{b}e_1...e_{s-2}})\,,
\end{align}
%$A^{kab}$ 
%a tensor constructed form the metric and antisymmetric in $(a,b)$
%\footnote{ we don't need detailed expressions but they can be found in \cite{Azeyanagi:2009wf}}.
while the boundary term in the variation of the action is \cite{Azeyanagi:2009wf}
\begin{align}\nonumber
\Theta=\frac12 & (-2\left( Z^{abcd}\nabla_d \delta g_{bc}-(\nabla_d Z^{abcd} \delta g_{bc}   \right ) \\ \label{bndrytermdef}
&\qquad \qquad+ \delta g_{ij} B^{ija} -Z^{alcd|e_1...e_{s-1} k}\delta \mathbb{R}^j_{kbcd|e_1...e_{s-1}}) )\epsilon_{abc} dx^b\wedge dx^c 
\,
\end{align}
with 
\begin{align}\nonumber
B^{ija}=&2 (Z^{ibcd |e_1...e_{s-1}a}+Z^{abcd |e_1...e_{s-1}i})\mathbb{R}^j_{bcd|e_1...e_{s-1}}
-2 Z^{ibcd|e_1...e_{s-1} j}\mathbb{R}^a_{bcd|e_1...e_{s-1}}\\ 
&+(s-1)(Z^{kbcd|e_1...e_{s-2} ia}\mathbb{R}^j_{kbcd|e_1...e_{s-2}i}
-\frac12 Z^{kbcd|e_1...e_{s-2} ij}\mathbb{R}^{\phantom{kbcd|}a}_{kbcd|\phantom{a}e_1...e_{s-2}i}) \label{corrB}\,.
\end{align}
%$B^{ija}$ a tensor constructed form the metric and symmetric in $(i,j)$.

The Iyer-Wald-entropy \cite{Iyer:1994ys} formula in 3D is 
\begin{equation}\label{Waldentropy}
\SWald=-2 \pi\int_{\text{horizon}} dA \,
 Z^{\alpha\beta\mu\nu}\epsilon_{\alpha\beta} \epsilon_{\mu\nu}\,
\end{equation} 
where $\epsilon_{\mu\nu}$ is the binormal at horizon and $dA$ the infinitesimal area element. 

We are interested in computing the charges associated to the exact Killing vectors $\partial_t$ and $\partial_\phi$. Their expressions are \cite{Wald:1993nt}
\begin{equation} \label{chargepf}
\delta L_0
%=\delta \mathcal{J}
\equiv-\int_\infty \delta Q_{\partial_\phi},
\end{equation}
\begin{equation} \label{chargept}
\delta P_0
%\equiv\delta \mathcal{E}
\equiv
\int_{\infty} \delta Q_{\partial_t} + \int_{\infty} i_{\partial_t}  \Theta \,,
\end{equation}
where the integral is taken on the circle at spatial infinity ($t,r=$cst) and $\delta$ is the derivation with respect to the parameters of the solution. There is no term with $i_{\partial \phi}\FTheta$ in $\delta L_0$ because $\partial_{\phi}$ is tangent to this circle.

%where $\xi_n$ are the asymptotic Killing vector, namely $\xi_n=-i \, n\, r e^{i\,n\,\phi}\partial_r+e^{i\,n\,\phi}\partial_\phi$ and 
%We are also interested in the computation of central charges. 
%We consider the case of a Virasoro algebra of conserved charges $L_m$ associated to asymptotic Killing vectors $l_n$ 
The asymptotic symmetries we are interested in are of three kinds: 
\begin{itemize}
\item two Virasoros for BTZ with Brown-Henneaux \cite{Brown:1986nw} boundary conditions
\begin{align} \nonumber
& i [L_m,L_n]=(m-n)L_{m+n}+\frac c{12}(m^3-m)\delta_{m+n,0}\, \\ \nonumber
&  i [\bar L_m,\bar L_n]=(m-n)\bar L_{m+n}+\frac {\bar c}{12}(m^3-m)\delta_{m+n,0}\, \\ \label{VirxVir}
&  i [L_m,\bar L_n]=0\,,
\end{align}
\item $Vir\ltimes u(1)$ algebra for BTZ with Comp\`ere-Song-Strominger \cite{Compere:2013bya} boundary conditions, here written in the quadratic ensemble \cite{Detournay:2012pc}
\begin{align}\nonumber
& i [\tilde L_m,\tilde L_n]=(m-n)\tilde L_{m+n}+\frac c{12}(m^3-m)\delta_{m+n,0}\\ \nonumber
& i [\tilde L_m,\tilde P_n]=-n\tilde P_{m+n}+n \tilde P_0\delta_{n+m}\\\label{Virxu1}
& i [\tilde P_m,\tilde P_n]=- 2m \tilde P_0  \delta_{m+n}\, ,
\end{align} 
\item BMS$_3$ \cite{Barnich:2006av} for flat space cosmologies
\begin{align}\nonumber
& i[L_m,L_n]=(m-n)L_{n+m}+c_{LL}\, m(m^2-1) \delta_{m,-n}\\\nonumber
& i[L_m,M_n]=(m-n) M_{n+m}+c_{LM}\,m(m^2-1)  \delta_{m,-n} \\ \label{bms3}
& i[M_m,M_n]=0\,.
\end{align}
\end{itemize}
The central extension of the algebra satisfied by the charges associated to the symmetry generators $\chi,\xi$ is given through the conserved $n-2$-form $k$ (see \cite{Barnich:2001jy,Barnich:2007bf} and references therein)
\begin{equation}
\int_{\Sigma} k_{\chi }[\delta_{\xi}\Phi, \Phi]
\end{equation}
where $\Phi$ are the fields of the theory and $\Sigma$ is a Cauchy surface.
%and $k$ is the conserved $n-2$-form (detailed expressions can be found for example in \cite{Azeyanagi:2009wf}). 
In the case of gravity, we have $\delta_\xi \Phi= \mathsterling_\xi \Phi$ and 
$k$ can be written as \cite{Lee:1990nz,Wald:1993nt,Iyer:1994ys} (for more details see \cite{Azeyanagi:2009wf,Detournay:2016gao}) 
\begin{equation} \label{k}
\int_\Sigma k_\chi[\mathsterling_\xi \Phi; \Phi ]
=\int_{\Sigma} \left (\delta_\xi Q_\chi(\Phi)+i_\chi \Theta(\mathsterling_\xi \Phi, \Phi)
\right ) \,.
\end{equation}
%\int_{\Sigma}-\left( \mathsterling_\xi  Q_\zeta -\mathsterling_\zeta Q_{\xi} \right)- i_\xi i_\zeta L -E \,
%\end{equation}
%Then, the central extension
%\footnote{When in the algebra, for example as in $Vir\times u(1)$, has a commutator of the form $ i [L_n,L_{-n}]=\frac c{\alpha}n\delta_{m+n,0}$, we use the same expression but we pick the term proportional to $n$.}
 %is given by
%\begin{equation}
 %c =\left .\alpha\, i \int_\Sigma k_{l_n}[\mathsterling_{l_{-n}}\phi,\bar \phi]\right|_{n^3} \,.
%\end{equation}
%where $k$ is the conserved $n-2$-form whom the expression is 
%\begin{equation} 
%\int_\Sigma k_\xi[\mathsterling_\zeta \phi; \bar\phi ]= 
%\int_{\Sigma}-\left( \mathsterling_\xi  Q_\zeta -\mathsterling_\zeta Q_{\xi} \right)- i_\xi i_\zeta L -E \,
%\end{equation}
%=\int_{\Sigma} \delta_\xi Q_\chi+i_\chi \Theta(\mathsterling_\xi \Phi, \Phi)
%In the literature, they are different definitions of the conserved $n-2$-form.
In the literature,  another definition for the $n-2$-form is made by Barnich-Brandt-Comp\`ere \cite{Barnich:2001jy,Barnich:2007bf}. Those two differ %from the Lee-Wald \cite{Lee:1990nz} formulation 
by the so-called $E-$term \cite{Azeyanagi:2009wf}
\begin{equation}
E[\mathsterling_{\xi_1}g;\mathsterling_{\xi_2}g]=\frac12\left(-\frac32 Z^{abcd}\mathsterling_{\xi_1}g_c^e\wedge \mathsterling_{\xi_2} g_{ed}+2Z^{acde}\mathsterling_{\xi_1}g_{cd}\wedge \mathsterling_{\xi_2} g^b_e \right)\epsilon_{abc} dx^c\,.
\end{equation} 
%The $E-$term encodes the difference between the approach of Iyer-Wald \cite{Wald:1993nt,Iyer:1994ys} and the one of Barnich-Brandt \cite{Barnich:2001jy}.
% Here, for all cases, it will be zero and so the two approaches give the same charges. 
 It turns out that in the considered cases, this term always vanishes and so the two definitions are equivalent. 
%Using the symmetries the fields $Z^{abcd},A^{kab},B^{ijk}$ are highly constrained and 
%The main advantage of that formalism is 
%that by the symmetries. 

\section{BTZ Black Holes}

\subsection{Symmetries}
%In this section, we constrain the general form of a tensor build from the metric. 
BTZ black holes are locally AdS$_3$. 
The latter is a maximally symmetric space. One main property of these spacetimes is that all tensors made out of the curvature and its covariant derivatives, like $R_{\mu\nu},\nabla_\mu R_{\nu\rho},...$ can be expressed in terms of the metric tensor. 
For example,
\begin{equation}
R_{\mu\nu}=-\frac{2}{\ell^2} g_{\mu\nu}
\end{equation}
where $\ell$ is the AdS radius. 
% We work with AdS$_3$ All the results of this section apply to BTZ because all the constrains we found here are derived locally.} 
Note that there is no way to construct a 3-tensor just from products of the metric. 
%Hence, any 3-tensor built from the metric and its derivatives has to vanish. 

Another consequence of the symmetries is that the left hand side of the equations of motion $E_{\mu\nu}$ evaluated on BTZ black holes, can always be rewritten as a constant times the metric
\begin{equation}
E_{\mu\nu}=K(\lambda_i,\ell) \, g_{\mu\nu} 
\end{equation}
where $K$ is a linear function of the coupling constants of the theory $\lambda_i$ as well as $\ell$ (but not on the parameters of the black holes). 
Solving the equations of motion imposes the constant $K$ to be zero which is in general solvable.
%Therefore, BTZ will be solution of any higher curvature gravity theory. 
%t is sufficient to hav leading to solve one equation for the coupling constant and $\ell$ the radius of AdS$_3$. In order to have a solution we only need one coupling constant. 

\subsection{Charges and Entropy}
Now, we can easily write the general form of the key field for the charges computation \eqref{Zabcd}: 
\begin{equation}
Z^{abcd}=\frac{\alpha}{32\pi} (g^{ac}g^{bd}-g^{ad}g^{bc}).
\end{equation}
%Indeed, it depends only on scalar invariant so it has to be written in terms of the metric and has to have the same symmetries as the Riemann. 
%It has the symmetries of the Riemann as it should. 
Indeed, as it depends only on curvature invariants, it can be written in terms of the metric and has the symmetries of the Riemann as it should. 
The constant $\alpha$ is a constant depending on $\ell$ and the coupling constants of the theory \footnote{The normalization is chosen for later convenience.}. 
Therefore, the Iyer-Wald entropy \eqref{Waldentropy} is proportional to the Bekenstein-Hawking entropy 
\begin{equation}
S^{IW}= \frac\pi 2 \alpha \, r_+=\alpha \,S^{BH}\,
\end{equation}
where $r_+$ is the outer horizon of BTZ. 

Moreover, the 3-tensors $A$ and $B$ in \eqref{Noetherchargesdef} and \eqref{bndrytermdef} vanish. The last term in \eqref{bndrytermdef}  can easily be shown to be zero as well (it is the variation of products of the metric, so it can be rewritten as a 3-tensor times $\delta g_{\mu\nu}$). 
%Another way to see their cancellation is to look at their definition, they are proportional to derivative of $\mathbb{R}_{abcd}$ who is zero for BTZ black holes. 
Thus, the expressions of the Noether charge \eqref{Noetherchargesdef} and the boundary term \eqref{bndrytermdef} are linear functions of the $Z^{abcd}$ field.
In addition, the higher derivatives of the Riemann in the Lagrangian do not have an effect on charges for BTZ black holes. Indeed, the terms of the sum with $i>0$ in \eqref{Zabcd} are always zero for the charges computation since they are the covariant derivatives of terms proportional to the metric.
%, Ricci tensor and derivatives of the Ricci which are zero on-shell. %evaluate $Z^{abcd}$ on-shell and as any tensor is p
Also, the higher derivatives might be present in the $i=0$ term as factor but will not contribute as they vanish on-shell.

% with coefficient independent of the considered theory.  
 %t is important to note that this field encodes all the information about the higher derivative theories. 
Therefore, the exact charges \eqref{chargepf} and \eqref{chargept} are proportional to their expressions in general relativity. All the information about the particular theory is encoded only through the proportionality constant $\alpha$ and there is no contribution from higher derivative terms of the Lagrangian, in the sense that they do not modify $\alpha$.

%As a consequence, the exact charges of BTZ black holes are the one of general relativity times the constant. 
%The corrections due to higher derivatives of the Riemann in Lagrangian can appear only through the $Z^{abcd}$. To be clearer we separate their contribution from non derivative case. 

Now, we compute the central charges. For that, we need to specify the boundary conditions.
Starting from the Brown-Henneaux boundary conditions \cite{Brown:1986nw}, we compute the central charges arising in the asymptotic symmetry group, generated by two copies of the Virasoro algebra. 
We use the following BTZ metric \cite{Banados:1992wn}
\begin{equation}
ds^2 =-\frac{\ell^2 r^2 dr^2}{-16 J^2 \ell^2+8 l^2 M r^2-r^4}+\frac{\left(8 \ell^2 M-r^2\right) dt^2}{\ell^2}+8 J dt \,d\phi+r^2 d\phi ^2 \,,
\end{equation}
where $J\leq\ell M$ and $\phi\sim \phi+2\pi$.
The conserved charges $L_n$ associated to the asymptotic Killing vectors
\begin{equation}
l_n=\left (\frac{\ell}{2}e^{i n \left(\frac{t}{l}+\phi \right)}+O\left(r^{-2} \right)\right )\pt
+\left (\frac{1}{2} i e^{i n \left(\frac{t}{l}+\phi \right)} n r+O\left(r^{-1} \right)\right )\pr
+\left ( \frac12 e^{i n \left(\frac{t}{l}+\phi \right)}+O\left(r^{-2} \right)\right )\pf \,
\end{equation}
satisfy one Virasoro of the algebra \eqref{VirxVir} \footnote{Since there is no gravitational anomaly, the two central charges are equal and we only need to consider one of the Virasoro's .}.
As the Noether charge \eqref{Noetherchargesdef} and the boundary term \eqref{bndrytermdef} are linear functions of $Z^{abcd}$, the central charge given by \eqref{k} is just proportional to the general relativity case
\begin{equation}
c=\alpha \left (\frac{3\ell}2 \right )\,.
\end{equation} 
As argued before, the exact charges are as well multiplied by the constant $\alpha$
\begin{equation}
M^{HC}=\alpha\, M \,,\quad J^{HC}=\alpha\, J \,.
\end{equation}
The dual theory is a CFT. The counting of the asymptotic density of states is given by the Cardy formula
%\begin{equation}
%S^{CFT}= 2\pi \sqrt{\frac c{12}(P_0 \ell -L_0)}+ 2\pi \sqrt{\frac c{12}(P_0 \ell+L_0)}= \frac\pi 2 \alpha \, r_+\,
%\end{equation}
\begin{equation}
S^{CFT}= 2\pi \sqrt{\frac c{6}L_0}+ 2\pi  \sqrt{\frac c{6}\bar L_0}\,,
\end{equation}
with $L_0=\frac{\ell \,M^{HC}+ J^{HC}}{2}$ and $\bar L_0=\frac{\ell \,M^{HC}- J^{HC}}{2}$. 
%with $P_0=Q_{\partial_t}=\alpha\,M $ and $L_0=Q_{\partial_\phi}=\alpha\,\,J$.  
So, 
\begin{equation}
S^{CFT}= \frac\pi 2 \alpha \, r_+\,,
\end{equation}
which is exactly the Iyer-Wald entropy.
%\begin{equation}
%S^{CFT}=S^{IW}=A\,  S^{BH}
%\end{equation}
%for any theory. 
We therefore have recovered the results of \cite{Saida:1999ec,Kraus:2005vz}.  

We continue with the Comp\`ere-Song-Strominger boundary conditions \cite{Compere:2013bya}. %The asymptotic symmetry group is  
%In that case, the dual field theory is believed to be a warped conformal field theory (WCFT) \cite{Detournay:2012pc}. 
%We worked in the coordinates used in \cite{Compere:2013aya} 
We first rewrite the BTZ metric as \cite{Compere:2013bya}:
\begin{equation}
ds^2 = \frac{\ell^2 dr^2}{r^2}+2 \ell (J+\ell M) dt_-^2+\frac{(4 J^2-4 \ell^2 M^2-\ell^2 r^4) dt_- dt_+}{r^2}-2 \ell (J-\ell M) dt_+^2 \,, 
\end{equation}
%with $J\leq \ell M$, $\phi\sim \phi+2\pi$ and $t_\pm=t/\ell\pm\phi$.
In these coordinates, the asymptotic Killing vectors are 
\begin{equation}
l_n=e^{i n t_+}\partial_{t_+} + \frac r2 i n e^{i n t_+}\partial_{r}\,, \quad t_n=e^{i n t_+}\partial_{t_-} 
\end{equation}
and their associated conserved charges $\tilde L_n,\tilde P_n$ satisfy the algebra \eqref{Virxu1}, i.e. 
%\cel{Problème pour vérifier le rhs de la deuxième équation} 
a $U(1)$ Kac-Moody-Virasoro algebra in the quadratic ensemble \cite{Detournay:2012pc}. 
%Once again, the $E-$term doesn't contribute and thus 
As argued before, the charges are proportional to their value in general relativity \cite{Compere:2013bya} 
\begin{equation}
c= \alpha\, \left (\frac{3\ell}2\right )
%\alpha \left (\frac{3\ell}2\right ) 
\, , \quad \tilde P_0=\alpha \, \left(\frac{ (J+\ell M )}2\right) \,,\quad
\tilde L_0= \alpha\, \left(\frac{ (-J+\ell M )}2\right) \,. 
\end{equation}
%In the dual description, these coordinates are the one of the quadratic ensemble presented in .
In that particular ensemble, the warped Cardy formula takes the form \cite{Detournay:2012pc}
\begin{equation}
S^{WCFT}=4\pi \sqrt{-\tilde P_0^{vac}\tilde P_0}+4\pi \sqrt{-\tilde L_0^{vac}\tilde L_0}
\end{equation}
where 
%$P_0$, $L_0$ are the zero modes of conserved charges associated respectively to $t_n$ and $l_n$ and 
$vac$ refers to the charges of the vacuum. Here, global AdS$_3$ is the vacuum whose charges are $M=-1/8$ and $J=0$.
%the vacua is given by taking $M=-1/8$ and $J=0$. 
So, we get 
\begin{equation}
S^{WCFT}=\frac\pi 2 \alpha \, r_+  \,. 
\end{equation}
Therefore, the gravitational entropy is also reproduced by the warped Cardy formula. 
%Once again, the higher derivative case reduces to the general relativity.
% one who were already proofed\cel{need citations}.

\subsection{Phase transition}
In this section, we consider the generalization of the Hawking-Page transition in the grand canonical ensemble for any theory of gravity where BTZ is a solution.
The exact charges are proportional to the general relativity case but the thermodynamic potentials, the temperature $T$ and the angular velocity $\Omega$ remain unchanged because their derivation relies only on the metric. 
Thus, the Gibbs free energy is given by 
\begin{equation}
 G^{HC}(T,\Omega)=\alpha\left ( \frac{ -\ell^2 \pi ^2 T^2}{2 \left(1-\ell^2 \Omega ^2\right)}\right )\,.
 \end{equation}

First, we study the local stability of that phase. In the grand canonical ensemble, the stability condition is the requirement for a system to have a negative semi-definite Hessian of its free energy $G(T,\Omega)$, given by
\begin{equation}
H=\begin{pmatrix}
\frac{\partial^2 G}{\partial T^2} & \frac{\partial^2 G}{\partial T \partial \Omega} \\
\frac{\partial^2 G}{\partial \Omega \partial T} &\frac{\partial^2 G}{\partial \Omega^2}
\end{pmatrix}
=
\alpha \left(
\begin{array}{cc}
-\frac{\pi ^2 \ell^2}{1-\ell^2\Omega ^2} & -\frac{2\ell^4 \pi ^2 T \Omega}{(1-\ell^2\Omega^2)^2} \\
- \frac{2\ell^4 \pi ^2 T \Omega}{(1-\ell^2\Omega^2)^2} & -\frac{\ell^4\pi ^2 T^2 (1+3\ell^2\Omega^2)}{(1-\ell^2 \Omega ^2)^3}
\end{array}
\right)\,.
\end{equation}
This implies only the following constraint on the coupling constants of the theory
\begin{equation}\label{localstabBTZ}
\alpha>0\,.
\end{equation}
% It is consistent with the fact that $P_0$ is asked to be bounded from below in the dual description. Therefore, we conclude that warped black holes are locally stable. 
%For example, the condition is satisfied because $\alpha= 1$. 

Secondly, we consider the global stability.
In the classical limit, the dominant phase is the most probable, i.e. the one that dominates the partition function among the saddle points. 
Here the two known phases are the black hole and the vacuum AdS$_3$. 
%The latter can be obtained through the BTZ metric by taking $M=-1/8,J=0$. Its free energy is $G^{HC}=\alpha \left(-1/8 \right)$. 
 So, we compare their free energies through their difference
\begin{equation}
\Delta G =\alpha \left(-\frac{1}{8}+\frac{\ell^2\text{  }\pi ^2 T^2}{ 2\left(1-\ell^2 \Omega ^2\right)}\right)\,. 
\end{equation}
If $\Delta G<0$, AdS$_3$ dominates and for the opposite sign, BTZ dominates.
The constant $\alpha$ factorizes out \footnote{We consider $\alpha>0$ because the study of the global stability at thermodynamic equilibrium requires to compare locally stable phase.}. It implies that the phase diagram doesn't depend on which theory we look at and it is the same as in general relativity.

\subsection{A working example: $L=a \,R-\Lambda+ b\, R^2 + c\, R_{ab}R^{ab}+d  \,R^3$}\label{sect.example}
We explicit the constant $\alpha$ of an example for which particular cases are general relativity and New Massive Gravity \cite{Bergshoeff:2009hq}.
%and $\alpha$ entering in \eqref{Zabcd} and the proportionality constant between the charges in higher curvature and their expressions in general relativity, $\alpha$. 
The auxiliary field $Z^{abcd}$ is 
\begin{align} \nonumber
Z^{abcd} 
&=\frac{\delta L}{\delta R_{abcd}}\\ \nonumber
&= (a +2  b  R+3 d R^2)\frac{\delta R}{\delta R_{abcd}} 
+2c R^{ef}\frac{\delta R_{ef}}{\delta R_{abcd}} \\ \label{example-general-Z}
&=\frac12 (a +2  b  R+3 d R^2)\left(g^{bd}g^{ac}-g^{ad}g^{bc} \right) 
+\frac12 c \left(g^{bd}R^{ac}-g^{ad}R^{bc}-g^{bc}R^{ad}+g^{ac}R^{bd} \right) \,.
\end{align}
The BTZ solution has $R=-\frac6{\ell^2}$ and $R_{ab}=-\frac2{\ell^2}g_{ab}$. So we get
\begin{equation}
Z^{abcd} 
=\frac{\alpha}{32\pi}\left(g^{bd}g^{ac}-g^{ad}g^{bc} \right) \,,\text{ with }\, \alpha=16\pi\left (a -\frac{12}{\ell^2}  b  + \frac{108}{\ell^4} d-\frac{4}{\ell^2}c \right )\,.
\end{equation}
Examples: 
\begin{enumerate}
\item The Lagrangian of general relativity is $\frac1{16\pi}R$. So the constant in $Z^{abcd}$ field is 
\begin{equation}
\alpha=1\,. 
\end{equation}
\item New Massive Gravity is obtained by taking $a=\frac1 {16\pi}$,$b=\frac{-3}{16\pi 8m^2 }$, $c=\frac1{16\pi m^2}$ and $d=0$ \cite{Bergshoeff:2009hq}, so 
\begin{equation}
\alpha=\left(1+\frac{1}{2 \ell^2 m^2 }\right)\,.
\end{equation} 
\end{enumerate}

\section{Flat Space Cosmologies}
\subsection{Charges and Entropy}
Now we turn to the case of flat space cosmologies (see \cite{Barnich:2012aw,Cornalba:2002fi} and references therein)
\begin{equation}
ds^2 = -2 dr du+8 M du^2+8 J du d\phi+r^2 d\phi^2
\end{equation}
with $M>0$ and $J\neq 0$. The cosmological horizon is located in $r_c=\sqrt{2J^2/M}$.
The asymptotic Killing vectors associated to the boundary conditions described in \cite{Barnich:2006av} are
\begin{equation}
l_n=e^{i n \phi }i n u \partial_u -i n r\pr\left(1+n^2\frac{u}{r}\right)\pf\,,\quad 
m_n=i e^{i n \phi }\partial_u\,.
\end{equation}
The associated charges $L_n,M_n$ satisfy the BMS$_3$ algebra \eqref{bms3}.
%\begin{align}\nonumber
%& i[L_m,L_n]=(m-n)L_{n+m}+c_{LL}\, m(m^2-1) \delta_{m,-n}\\\nonumber
%& i[L_m,M_n]=(m-n) M_{n+m}+c_{LM}\,m(m^2-1)  \delta_{m,-n} \\
%& i[M_m,M_n]=0\,.
%\end{align}

We show that the charges are always the same regardless of the theory. 
%We recover always the general relativity charges. 
%We could have done the similar analysis with symmetries. Here, it easier to work directly with the expressions of charges. 
Indeed, in these solutions, $R=R_{\mu\nu}=0$ and covariant derivatives of the Ricci all vanish, 
% and the charges computation is an on-shell procedure. 
fixing completely $Z^{abcd}$ \eqref{Zabcd}. 
Its term $i=0$ 
%takes into account the power 
takes the derivatives with respect to the Ricci tensors. To have a non zero contribution, we need at least and at most one Ricci tensor, because if it is multiplied by any ingredient of the Lagrangian apart from the metric, it will no contribute on-shell. 
%The latter can be multiplied by any other ingredients of the Lagrangian $g_{\mu\nu},R,R_{\mu\nu},\nabla_{e_1}...\nabla_{e_n} R_{\mu\nu},...$. But only the term $g_{\mu\nu}R^{\mu\nu}$ will give a non zero contribution on-shell. 
%because all the others are proportional to something vanishing on-shell. 
%When there more than $2$ Riccis \footnote{we mean any objects with a Ricci inside: $R,R_{\munu},\nabla_{e_1}...\nabla_{e_n} R_{\mu\nu}$, }, its derivative brings a factor proportional to the Ricci and it cancels on-shell. 
The terms $i>0$ are always zero by the same argument as for BTZ. 
%Therefore, as we will see in details, only the term linear in the Ricci will have an non zero contribution. 
%First we consider the expression of $Z^{abcd}$ \eqref{Zabcd}. 
%The term of the sum with $i>0$ will be zero by the same argument as for BTZ. 
%The only not zero contribution in the term $i=0$ is $ a R$ in the Lagrangian. Indeed, all the higher curvature term will be proportional to $R_{\mu\nu}$ or $R$ and so will be zero on-shell.
So in general, the $Z^{abcd}$ field is of the form
\begin{equation}
Z^{abcd}=\frac1{32\pi}\left(g^{bd}g^{ac}-g^{ad}g^{bc} \right) \,. 
\end{equation}
For example, if we take the Lagrangian of Sect. \ref{sect.example}, the steps are identical to BTZ until the Eq. \eqref{example-general-Z}. On-shell, only the term proportional to $a$ survives. It is the constant of the Einstein-Hilbert term which is usually taken to be $1/(16\pi)$. 
\footnote{If we consider a theory with neither an Einstein-Hilbert or Chern-Simons term having flat spaces cosmologies as solutions, their charges will be zero.}

Therefore, the Iyer-Wald entropy is just the general relativity expression, namely 
\begin{equation}
S^{IW}=\frac{\pi}{2}r_c \,.
\end{equation}
Furthermore, the corrections tensors $A$ \eqref{corrA} and $B$ \eqref{corrB} are proportional to $\mathbb{R}_{abcd|e_1...e_s}$ \eqref{auxRabcd} which is always zero on-shell.
%\footnote{A priori, $Z^{abcd|e_1...e_s}$ can be non zero. The terms with only one power of the Ricci or one covariant derivative of the Ricci will contribute. To have such term in the Lagrangian, we only can have a even number of covariant derivatives in order to be able to build a scalar.
% In that case $Z^{abcd|e_1...e_s}$ will be proportional to products of the metrics for $s$ even and zero for $s$ odd. }.
%Also, the $E-$term vanishes. 
Thus, we recover the general relativity result \cite{Barnich:2012aw}
\begin{equation}
Q_{\partial_u}=M\,,\qquad Q_{\partial_\phi}=J
\end{equation}
\begin{equation}
c_{LL}=0\,,\qquad c_{LM}=\frac14 \,. 
\end{equation}
%All the result about charges of \cite{Barnich:2012xq},\cite{Bagchi:2012xr} hold in any higher curvature theories. 
%Another way to obtain these results is by considering dimensional analysis. 
This result is consistent with the argument following from dimensional analysis.
The symmetries of these spacetimes imply that the charges in any higher curvature theory are proportional to their value in general relativity. The proportionality constant is only a linear combination of the coupling constants of the theory whose the coefficients are pure numbers as locally flat spacetimes do not have an intrinsic length. Moreover, this combination should be consistent with dimensional analysis. Thus, the only allowed term is the Einstein-Hilbert term and in the case of theory with diffeomorphism anomaly, also the Chern-Simons term.

One consequence of the non renormalization of the charges is that the results about charges of \cite{Barnich:2012xq,Bagchi:2012xr} hold in any higher curvature theories.
For example, their entropy is reproduced by the Cardy-like formula for BMS$_3$ \cite{Barnich:2012xq,Bagchi:2012xr}.

\subsection{Phase transition}
In \cite{Bagchi:2013lma},  a phase transition between flat spaces cosmologies and hot flat spaces is exhibited. 
Nevertheless, our local stability analysis differs from theirs and modifies the conclusion. 

We consider the flat space cosmologies written for convenience in the following form
 \begin{equation}\label{FSC}
 ds^2=R_+^2 dt^2 -2R_+r_0 dtd\phi - \frac1{R_+^2(1-r_0^2/r^2)}dr^2+r^2d\phi^2\,.
 \end{equation}
 The temperature and angular velocity are \cite{Bagchi:2013lma}
\begin{equation}
T=\frac{R_+^2}{2\pi r_0}\,, \quad \Omega=\frac{R_+}{r_0}\,.
\end{equation}
We directly consider the case of TMG \cite{Deser:1981wh,Deser:1982vy}
\begin{equation}
  S=\frac1{16\pi}  \int d^3x\sqrt{-g}(R-2\Lambda)
  + \frac1{16\pi }\frac1{2\mu}  \int d^3x\sqrt{-g}\epsilon^{\lambda\mu\nu}\Gamma^\rho_{\lambda\sigma}\left(\partial_\mu\Gamma^\sigma _{\rho\nu}+ \frac23\Gamma^\sigma _{\mu\tau}\Gamma^\tau_{\nu\rho} \right)
\end{equation}
with $\mu$ the Chern-Simons coupling taken positive without loss of generality. 
The conserved charges associated with $\partial_t$ and $\partial_\phi$ are \cite{Bagchi:2013lma}
\begin{equation}
 M=\frac{R_+^2}8 \,,\quad  J=-\frac{R_+r_0}4+\frac{R_+^2}{8\mu}\,, 
\end{equation}
while the entropy is \cite{Bagchi:2013lma}
\begin{equation}
S=\frac{\pi r_0}{2}-\frac{\pi R_+}{2\mu}\,.
\end{equation}
To study the thermodynamic stability, we need to consider the Gibbs free energy $G$ who can be derived by integrating the first law $ dM=-TdS-\Omega \,dJ$ \cite{Bagchi:2013lma}, or through the on-shell action procedure \cite{Grumiller:2015xaa}, 
\begin{equation}
G=M+TS+\Omega J
\end{equation}
  leading to \footnote{The free energy has an opposite sign with respect to \cite{Bagchi:2013lma}, because here it is expressed in terms of Lorentzian variables.} %$^\text{,}$

  \begin{equation}\label{GFSC}
  G^{FSC}(T,\Omega)=\frac{\pi^2T^2}{2\Omega^2}\left (1-\frac{\Omega}{\mu} \right )\,.
  \end{equation}
%In \cite{Bagchi:2013lma}, the analysis was made in the grand canonical ensemble, where the free energies in given in 
%\begin{equation}
%G^{HFS}=-\frac18 \,, \qquad 
%G=-\frac{\pi^2T^2}{2\Omega^2}
%\end{equation}
%where $T=\frac{2 \sqrt{2} M^{3/2}}{J \pi }$ is the temperature and $\Omega=\frac{2 M}{J}$ the angular velocity. 
 In the grand canonical ensemble, it is not sufficient to study only the specific heat to determine the local stability as was done in \cite{Bagchi:2013lma}. 
%The free energy has to be a concave function of $T,\Omega$ in order to have a locally stably phase. 
Instead, the complete requirement is that the Gibbs free energy should be a concave function of the temperature and the angular velocity. In other words, its Hessian $H$ should be negative semi-definite which is never the case here
\footnote{It is well known that the flat cosmologies can be obtained as the flat limit of BTZ black holes. 
Nevertheless, the flat limit of the condition \eqref{localstabBTZ} obtained for BTZ does not give to the one for flat space cosmologies. Indeed, it is a consequence of the local stability condition in where $\alpha$ and the derivatives of the free energy take place. Therefore, we should take the limit of all these ingredients and after discuss. This procedure leads to the result presented in this section.
%Doing so, it leads in fact to the same conclusion as presented in this section.
}
. Indeed, it is given by
\begin{equation}
H=\begin{pmatrix}
 \frac{\partial^2 G}{\partial T^2} & \frac{\partial^2 G}{\partial T \partial \Omega} \\
\frac{\partial^2 G}{\partial \Omega \partial T} &\frac{\partial^2 G}{\partial \Omega^2}
\end{pmatrix}
=
\left(
\begin{array}{cc}
 \frac{\pi ^2 \left(1-\frac{\Omega }{\mu }\right)}{\Omega ^2} & -\frac{\pi ^2 T (2 \mu -\Omega )}{\mu  \Omega ^3} \\
 -\frac{\pi ^2 T (2 \mu -\Omega )}{\mu  \Omega ^3} & \frac{\pi ^2 T^2 (3 \mu -\Omega )}{\mu  \Omega ^4}
\end{array}
\right)\,.
\end{equation}
The determinant of $H$ is $-\frac{\pi ^4 T^2}{\Omega ^6}$. It is negative and the two eigenvalues values are non zero (expect in the limit $\mu\to0$ which will be considered below). It implies that the eigenvalues have not the same sign and so the matrix can not be negative semi-definite. 
So the flat spaces cosmologies are not metastable states and therefore the phase transition can not be studied in the frame of the physics of thermodynamic equilibrium. The extensions for non equilibrium phenomena are far beyond the scope of this paper. 
%therefore there is no point to study the global stability.
It is also the case of pure general relativity ($\mu\rightarrow\infty$) and therefore in any diffeomorphism invariant theories. 
%Study the global stability of a non locally stable phase is not well defined because global stability compares minimum of free energy \cel{bof}
%One important remark is the particular case of TMG. In that case, the charges are modified \cite{Bagchi:2013lma}.
% where is has been shown \cite{Bagchi:2013lma} that the constant $A$ takes the form 
%For example, the entropy is given by 
%\begin{equation}
%S^{TMG}= \frac{\pi}{2}\left (r_c +\frac1\mu \sqrt{8M}\right )\,. 
%\end{equation}
%In particular the free energy takes the form
%\begin{equation}
%G=-\frac{\pi^2T^2}{2\Omega^2} \left (1 +\frac\Omega\mu \right )\,.
%\end{equation}
%As the free energy is modified, we can reanalyse the local stability by the same procedure. In TMG, it is never stable but in Flat Chiral Gravity \cite{Bagchi:2012yk}, where only the Chern-Simons remains, it turns out that the Hessian is negative semi-definite and thus that flat space cosmologies are locally stable. 

Only in the particular case of flat chiral gravity \cite{Bagchi:2012yk}, where only the Chern-Simons remains, it turns out that the Hessian is negative semi-definite for positive angular velocities
\begin{equation}
\Omega>0\,.
\end{equation}
If we naively pursuit the analysis, we will conclude that the flat space cosmologies are locally stable and therefore we will consider the global stability
%Naively, we proceed to the usual thermodyanamic analysis but as we will explain below, we can not make sense of the thermodynamic for these solutions in that theory.
%But the thermodynamic ensemble for these spacetimes in that theory are maybe ill define because the eigenvalue of $M_0$ becomes zero. That generator was related to $r_+$ and ASG is one Virasoro (chiral). Maybe there is a trivial diffeo who can modify the position of $r_0$ and therefore    
%and thus the flat space cosmologies are locally stable. 
%Therefore, in that context we can analyse the global stability.
%\footnote{We will consider the phase diagram determined through the difference of free energies \eqref{GFSC}, \eqref{GHFS} in the domain where $\Omega>0$,
%\begin{equation}\Delta G=(G^{HFS}-G^{FSC})= \frac{1}{\mu\,\Omega }\left (\frac{\pi ^2 T^2}{2 }-\frac{\Omega^2 }{8}\right )\,,\end{equation}
%and conclude that the phase transition happens along the coexistence curve ($\Delta G=0$), $R_+=1$, at the critical temperature
%\begin{equation}\label{criticaltemp}T=\frac{\Omega }{2 \pi }=\frac1{2\pi r_0}\,,\end{equation}
%corresponding to the self-dual point of the S-transformation \cite{Bagchi:2013lma}.
%Above it, the flat space cosmologies dominate over the hot flat spaces and below, the contrary.}.
However, the limit $\mu\rightarrow 0$ leads to a theory with an enhancement of symmetry. Indeed, the action is a pure Chern-Simons term who is conformally invariant. Rewritting the flat space cosmologies \eqref{FSC} as
\begin{equation}
 ds^2=r_0^2\left (R_+^2 dt^2 -2R_+ dtd\phi - \frac1{R_+^2(1-1/r^2)}dr^2+r^2d\phi^2\right )\,,
\end{equation}
%through the transformation $t \rightarrow r_0\, t$ and $r \rightarrow r_0 \,r$,
 it is obvious that the value of $r_0$ can always be modified by a conformal transformation. 
Therefore we can not make sense of the thermodynamics of the flat space cosmologies in flat chiral gravity and obviously it is meaningless to talk about phase transition in that context.  

In conclusion, we exclude the possibility of a phase transition between flat spaces cosmologies and hot flat spaces for any theory of gravity, with or without gravitational anomalies, including in the analysis the singular point $\mu\to 0$. 
%We can ask ourself what can happen in the particular limit of NMG where $m^2 \rightarrow \infty$ where there is an enhanced symmetry of the form of a covariant symmetry. But in that case, as we saw before all the charges are zero and nothing happens.

\section{Conclusion}

%In this paper we have shown that in any generic higher curvature theory of gravity, the exact and central charges of BTZ black holes are given by a renormalization of general relativity charges and moreover that the presence of higher derivative of the Riemann in the Lagrangian doesn't affect the charges.
In this paper, we have analysed in detail the corrections due to higher derivative terms in the action and have shown that they never contribute to the renormalization of the charges.
Also, we have proved that the Iyer-Wald entropy of BTZ black holes, equipped with the Comp\`ere-Song-Strominger boundary conditions, is reproduced by the warped Cardy formula. 
As a direct consequence of the charge renormalization of BTZ black holes, the same Hawking-Page transition occurs as in general relativity since the constant of proportionality factorizes out. 
Nevertheless, the thermodynamic local stability of BTZ black holes imposes an additional constraint on the coupling constants of the theory. 

In addition, we have proved that the charges of flat space cosmologies are never modified in any diffeomorphism invariant theory. Therefore, the properties of the charges derived in general relativity in \cite{Barnich:2012xq,Bagchi:2012xr}, for example the match between the bulk and boundary entropies, holds in general. 
We have also discussed the phase transition between flat space cosmologies and hot flat spaces where we found a few discrepancies with the work of \cite{Bagchi:2013lma}. %Our results contradict the previous work \cite{Bagchi:2013lma}. 
We have argued that the local stability requirement excludes the possibility of a phase transition at thermodynamic equilibrium, except potentially in the case of flat chiral gravity. However, we have pointed out that the thermodynamics of flat space cosmologies can not be considered in flat chiral gravity because of its property of conformal invariance. 
%the global stability analysis leads to a phase transition. 

%We have also clarified the situation about the local stability in the grand canonical ensemble and excluded the possibility of a phase transition between the flat space cosmologies and hot flat spaces, except in the case of flat chiral gravity where a transition occurs. 
%where the results of \cite{Bagchi:2013lma} holds 
%we leave the question opened. 

\section*{Acknowledgements}
The author is grateful to S. Detournay for its comments on the draft and enlightening discussions. She also thanks G. Giribet, H. Gonz\'alez, D. Grumiller, V. Lekeu, A. Marzolla, and G. Ng. 
She is a research fellow of ``Fonds pour la Formation à la Recherche dans l'Industrie et dans l'Agriculture"-FRIA Belgium. This work is partially supported by the ARC grant ``Holography, Gauge Theories and Quantum Gravity - Building models of quantum black holes" and by FNRS-Belgium (convention IISN 4.4503.15).
%\bibliographystyle{utphys}
%\bibliographystyle{utphys}
%\bibliography{all.bib}

%\providecommand{\href}[2]{#2}\begingroup\raggedright\begin{thebibliography}{10}

%\endgroup

\end{document}